# A Multi-Agent System of Project Bidding Management Simulation

R. Liu

ABSTRACT

This paper presents a simulation model based on the general framework of Multi-Agent System (MAS) that can be used to investigate construction project bidding process. Specifically, it can be used to investigate different strategies in project bidding management from the general contractors' perspective. The effectiveness of the studied management strategies is evaluated by the quality, time and cost of bidding activities. As an implementation of MAS theory, this work is expected to test the suitability of MAS in studying construction management related problems.

INTRODUCTION

Project bidding is a multi-disciplinary and multi-organizational process that requires the efforts of different project functional units (Kerzner 2009). Unlike intra-team activities, project bidding happens in a cross-functional environment where a formal boundary between responsibilities is set and leads to diverse institutional arrangements (Thomsen et al. 2005). For example, in an EPC (Engineering, Procurement and Construction) project, ideally the estimating team and the engineering team work closely together on developing a proposal; but in reality, the two teams have distinct responsibilities, focuses and procedures. This often results in additional work such as coordination, and without sufficient management, rework is almost inevitable. Another difficulty is the bid/no bid decision. Some scholars have applied machine learning approaches (Du and El-Gafy 2011), statistical modeling, building information modeling (BIM)(Du et al. 2014; Liu et al. 2014; Liu et al. 2014) or Monte Carlo simulation (Du et al. 2014) to support a better decision, but the bid/no bod decision remain a challenge for the construction managers.

A root cause of the inefficiency in bidding management is the lack of understanding about the proper management strategies (Du and El-Gafy 2014). One example is the goal incongruence (Du and El-Gafy 2014): the estimating team may make the economy of the proposed design its first priority, while for the engineering team, robustness is more important. Such difference in perception may lead to completely different practices. In order to improve the efficiency and effectiveness of bidding process, it is critical for general contractors to understand the consequences of different management strategies and approaches. It involves the optimization of number of target projects, the job assignment strategies and meetings.

Many existing efforts concentrate on only one aspect of human behaviors pertaining to bidding management, assuming that a deeper investigation on a single aspect will lead to better discovery. The rationale of focusing on one important point is well recognized by this study, especially given the difficulties of conceptualizing human behaviors and validating assumptions. Nonetheless, the importance of addressing as many relevant behaviors as possible in the same investigation should not be intentionally overlooked, when the interactions among diverse behaviors play a critical role in understanding how goals are formed and affected and how goal incongruence influences the efficacy and quality of proposal development (Perrow 1986).

This paper introduces a simulation model based on Multi-Agent System (MAS) framework, to investigate the implications of management strategies in the bidding management of a small construction project. Worker behaviors pertaining to bidding were captured and investigated to quantify the impacts of different management strategies on the performance of bidding management of a general contractor.

LITERATURE REVIEW

As a computational modeling approach, Agent Based Modeling (ABM) is a suitable tool for use in social research to study human and organizational issues in a diversity of areas (Du 2012; Du and El-Gafy

2010; Du and Wang 2011). It is a computational method that builds a common environment for heterogeneous and autonomous agents to share, and allows the agents to simultaneously interact with each other for self-interest (Du and El-Gafy 2014; Ligmann-Zielinska and Jankowski 2007). Unlike top-down modeling approaches (e.g., System Dynamics, Discrete Event Simulation etc.), in ABM the collective behavior of the simulated system is not predefined, but emerges from individual agents who act based on what they perceive to be their own interests. Thus, ABM is capable of reproducing the emergent properties of the studied systems (Macal and North 2007).

As for the application of ABM in construction engineering and management, recently Du and colleagues have performed a series of representative works (Du 2014; Du and Bormann 2014; Du and El-Gafy 2010; Du and El-Gafy 2012; Du and El-Gafy 2014; Du and El-Gafy 2014; Du et al. 2012; Du et al. 2014; Du and Wang 2011). In one of their works, they developed a comprehensive ABM model called "Virtual Organizational Imitation for Construction Enterprises" or "VOICE". In the VOICE model, they creatively captured 13 common behaviors in construction management settings, and simulated them under the MAS framework. Unlike other similar works, in the VOICE model, a comprehensive list of work related behaviors are modeled as separate behavioral modules. It suggests a better capture of the sociotechnical process of construction management. Given the features of VOICE, this study mainly builds its simulation experiments on VOICE model.

THE MODEL ARCHITECTURE
In order to utilize the VOICE framework to investigate problems discussed above, the following basic assumptions were made:

First, there are three major agents including president, who is responsible for the overall management and decision-making of bidding; managers who are responsible for information gathering and expectation handling; and helpers who are responsible for processing routine tasks. Figure 1 illustrates the MAS used in the modeling and simulation.

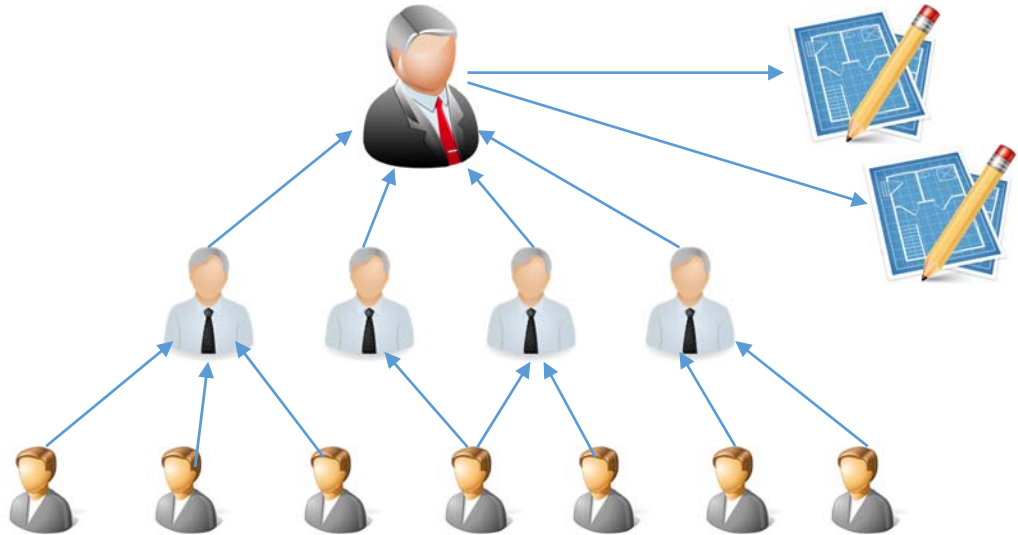

Fig.1 MAS architecture of the proposed model

Second, the decisions made by the agents may trigger a variety of individual behavioral responses modeled with the behaviors in the VOICE framework. Typical behaviors include routine activities (e.g., processing tasks), communication, and coordination (e.g., assigning tasks). However, when overloaded, reciprocal activities may also be triggered, such as complaining about the overload. These nonproductive activities create inefficiency and affect the capacity of the estimating team.

Third, although under the VOICE framework, task characteristics and organizational context can also affect the cooperative behaviors of team members, they will not be considered in this case study because they are less dynamic in Company D, compared to the four issues addressed by the principals. Therefore, the simulation experiments only focused on the controllable variables for a realistic recommendation.

Based on the basic assumptions, the proposed modeling framework is shown in Figure 2.

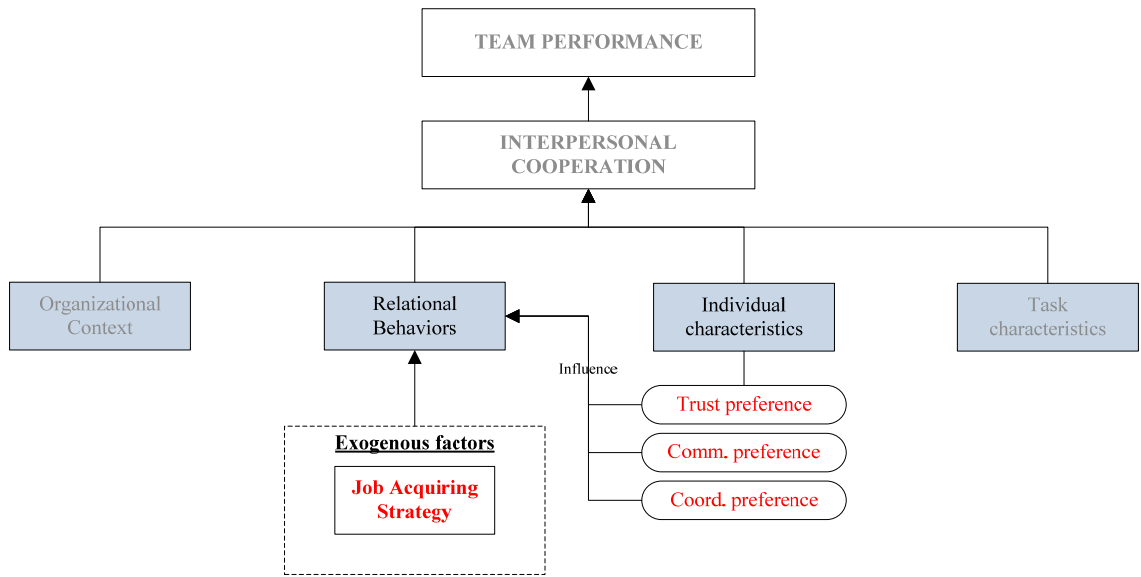

Fig. 2 Modeling bidding management with proposed theoretical framework

## CASE STUDY

A case study was performed to investigate how behaviors and the institutional arrangement between members of a single project team affect management actions and team performance in typical Design Bid Build (DBB) projects. The studied case is a small general contractor focused on small commercial projects. Most jobs are Design Bid Build (DBB). The cost estimation is conducted by a single team: three managers work on separate sections/crafts of the project and all report to the principals for the final estimating and bidding decision. In the simulation experiments, it is of particular interest to test:

- **The influence of task dependency:** Among all the task-related factors, dependence among tasks is considered to be most correlated to the level of cooperation (Deutsch 1949; Pinto et al. 1993; Thompson 2003). Task dependence refers to the extent to which team members are dependent on each other to perform individual tasks (Van de Ven et al. 1976). The original work about task dependence can be dated back to Thompson (1967), who grouped task dependence into three types -- pooled, sequential and reciprocal -- with reciprocal dependence at their highest intensity of interaction. Regarding construction as a complex system (Bertelsen 2003), reciprocal task dependence is probably the most common dependence in construction project teams (Thompson 2003). Because reciprocal task dependence means the highest level of interaction intensity (Thompson 1967), intense coordination work is required to adjust the efforts of different actors (Levitt 2007). Building upon Thompson, it was induced that the hierarchy of increasing levels of task dependence between unit personnel can be determined by observing whether the work flow is (1) independent, (2) sequential, (3) reciprocal, or (4) in a team arrangement (Van de Ven et al. 1976). Yilmaz and Hunt (2001) proposed measuring task dependence by the information need of tasks, i.e., whether additional information is needed to perform a particular task. Following the previous work, this research describes task dependence in construction project teams as the workflow relationship between team members, which can be demonstrated by network techniques, such as activity on node (AON).

- <u>The influence of goal congruence.</u> In the bidding management process goal congruence plays a vital role in this process, which is demonstrated in the difference of the perceptions of behavioral standards and ranking of management criteria (Thomsen et al. 2005). Goal congruence can affect the quality and amount of the appropriate information contributed by the designers because a higher magnitude of goal congruence is anticipated to enhance the understanding among team members (Witt 1998). Thomsen et al. (2005) model goal congruence as a percentage, with 100% being the most congruent condition and 0% being the least. This case study uses the same definition and assumes a linear relationship between goal congruence and information quality/amount exchanged between an engineer and a project proposal team member.

## Simulation results

3,300 simulations were conducted to examine the influence of task dependence and goal congruence on the performance of the bidding team. The following figure demonstrates the results.

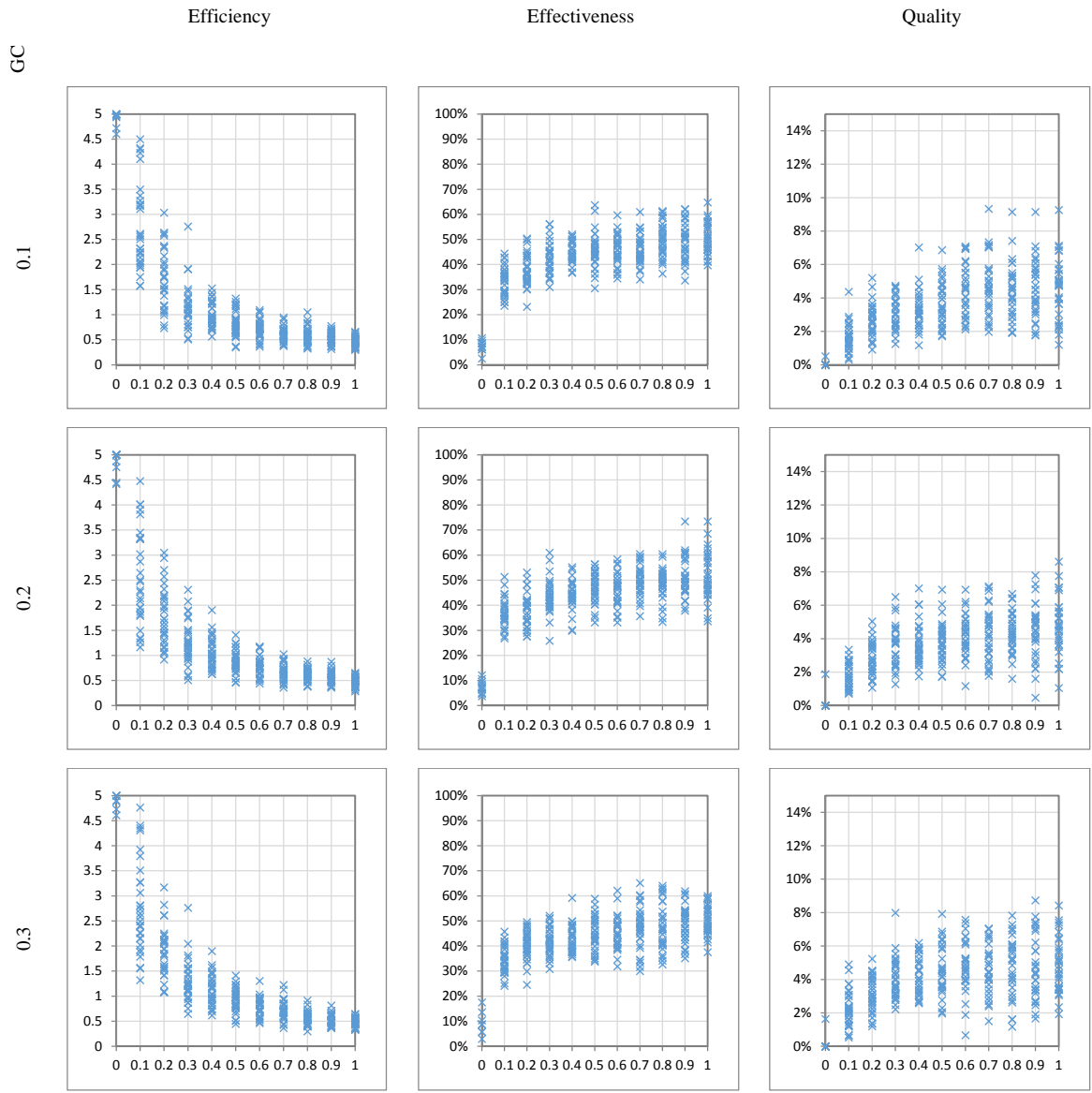

0.4

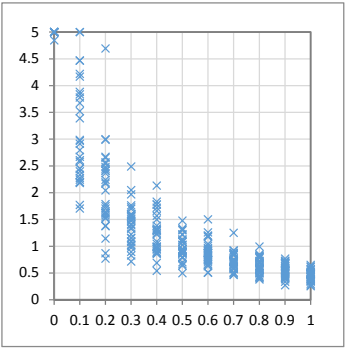 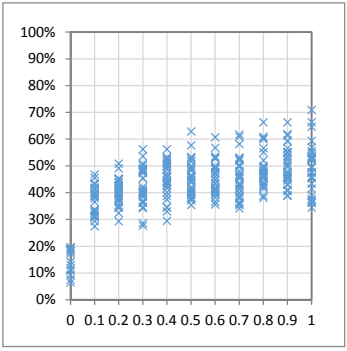 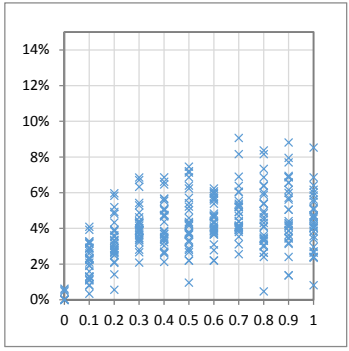

0.5

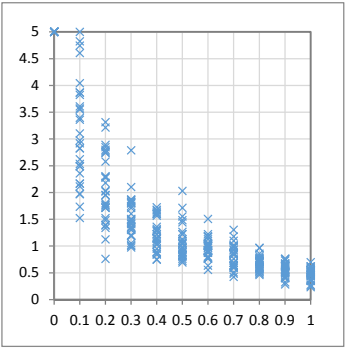 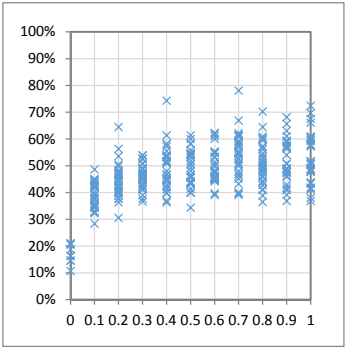 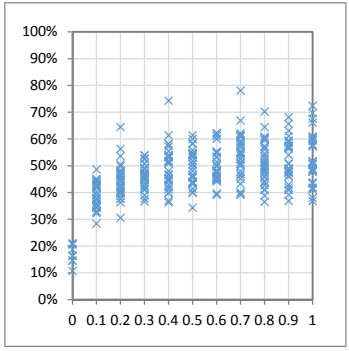

0.6

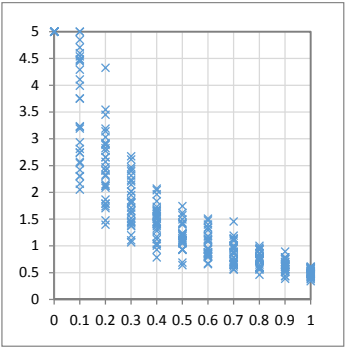 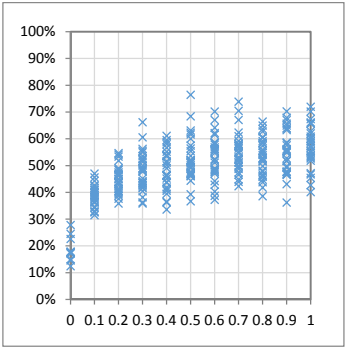 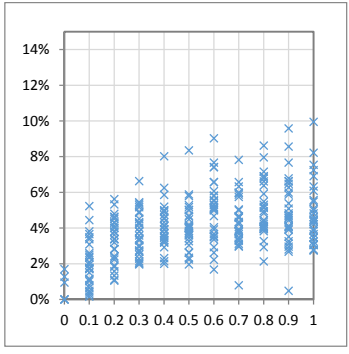

0.7

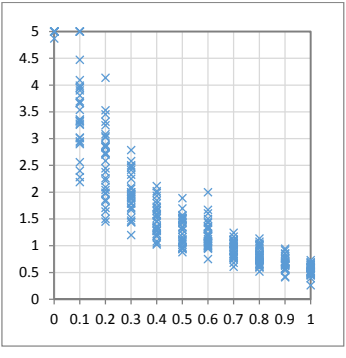 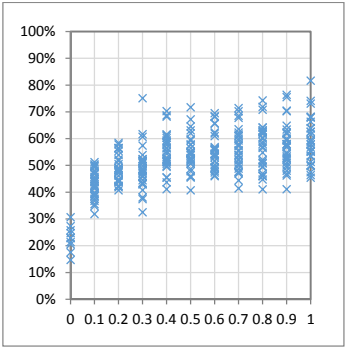 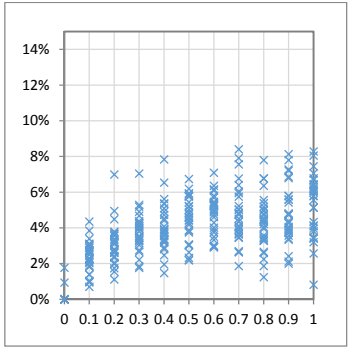

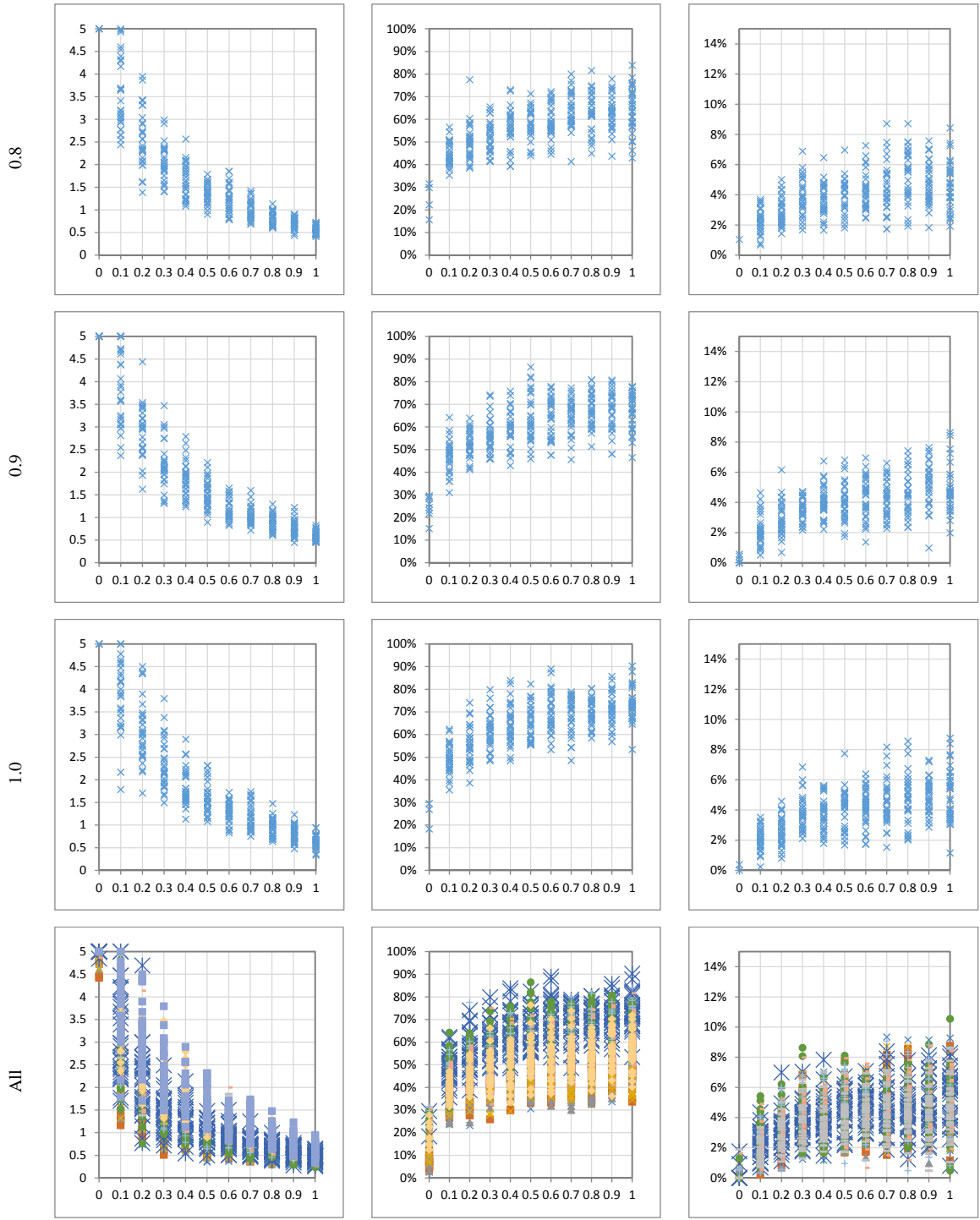

**Fig. 3 Influences of task dependence and goal congruence on performance of proposal development**

## DISCUSSION AND CONCLUSIONS

The case study made two major findings. First, the simulation finds goal congruence to be an influential factor for team productivity, but negligible to the work quality and work pressure of the project team.

First, a higher level of goal congruence between the proposal team and engineering team significantly improves the efficiency and effectiveness of proposal development. A likely interpretation is that enhanced goal congruence improves the mutual understanding of objectives, definitions and needs between two teams, and encourages proactive participation of the engineers in proposal development. This in turn reduces the need for additional coordination, and increases the quality of each information exchange between engineers and proposal team members. Second, task dependence can significantly affect the productivity and work quality of the project team; it is able to alter the effects of goal congruence and micro-management. Task dependence is a crucial factor for understanding inter-team cooperation in proposal development. On the one hand, the simulation results find task dependence to be a significant predictor of efficiency, effectiveness, and quality. If tasks are more dependent, the team is less productive and commits more mistakes. This is understandable from an empirical perspective, since dependence often means additional efforts for communication and coordination, a bigger chance of mistakes and conflicts. On the other hand, task dependence may affect the effects of goal congruence and micro-management. Simulation results found that the efficiency difference between levels of goal congruence becomes bigger when tasks are more dependent. In contrast, the effects of micro-management are more significant when tasks are more independent. This finding highlights task dependence to be a vital point of decision making in project team management, especially when managerial and/or behavioral changes are planned.